\documentclass[apj]{emulateapj}
\slugcomment{Accepted for publication in ApJ}

\usepackage{paralist}
\usepackage{graphics,graphicx}

\shorttitle{Alternating lags in QPO}
\shortauthors{Misra \& Mandal}

\def\mathnew{\mathsurround=0pt}

\def\simov#1#2{\lower .5pt\vbox{\baselineskip0pt \lineskip-.5pt
\ialign{$\mathnew#1\hfil##\hfil$\crcr#2\crcr\sim\crcr}}}

\begin{document}

\title{Alternating lags of QPO harmonics $:$ A Generic model and its 
application to the 67 millihertz QPO of GRS 1915+105}

\author{Ranjeev Misra\altaffilmark{1} and Soma Mandal\altaffilmark{2}}

\altaffiltext{1}{Inter-University Center for Astronomy and
Astrophysics,  Post Bag 4, Ganeshkhind, Pune-411007, India;
rmisra@iucaa.ernet.in}

\altaffiltext{2}{Department Of Physics, Taki Government College, Taki,
North 24 Parganas-743429, West Bengal, India; soma@iucaa.ernet.in}

\begin{abstract}
A generic model for alternating lags in QPO harmonics is presented
where variations in the photon spectrum are caused by oscillations
in two parameters that characterize the spectrum. It is further assumed that
variations in one of the parameters is linearly driven by variations
in the other after a 
time delay $t_d$. It is shown that alternating lags will be observed 
for a range of $t_d$ values. A phenomenological model based on this
generic one is developed which can explain the
amplitude and phase lag variation with energy of the fundamental
and the next three harmonics of the $67$ mHz QPO observed in GRS 1915+105. 
The phenomenological model
also predicts the variation of the Bicoherence phase with energy,
which can be checked by further analysis of the observational data.

\end{abstract}

\keywords{accretion, accretion disks-black hole physics -stars: individual (GRS 1915+105)}

\section{INTRODUCTION} \label{sec: I} 

Quasi-periodic oscillations (QPO) have been observed in both black
hole and neutron star systems. For black hole systems QPO have
been observed for a wide range of frequencies from mHz to
a few hundred Hz \citep{MR06,RM06,V06},
 while kHz QPO have also been observed in neutron star systems
\citep{V00,V06}. The high
frequency of these oscillations suggests that they probably originate
in the innermost regions of these system and hence in principle could
be used to probe and test effects of general  relativity in the strong field limit. Since some
of the low frequency QPO are observed to be correlated to the high frequency
ones, they may have a common origin. Moreover, a study of the 
temporal behavior of
these systems (particularly QPO activity) could give new insight into
the enigmatic nature of these systems.

Theoretical studies of the QPO phenomena have generally been
restricted to identifying the different frequencies as natural
characteristic frequencies of the system and correlation between
them \cite[e.g.][]{Mi98,SV99,T099}.
A consistent model incorporating radiative
and dynamic mechanisms, that translates the natural frequency
of the system to an observable QPO, remains illusive. Clues to
such underlying processes can be obtained by studying the energy
dependence of the QPO features like amplitude and phase lags 
 \citep[e.g.][]{Cui99,LMT,Li13,Lin,M97,QU10,Reig,RJ02}, especially
when harmonics are observed since additional information can be obtained for
the same phenomenon. Such an endeavor has been impeded, because the
energy dependence of the phase
lags for many QPO are observed to be complex and contrary to expectations \citep{Cui00, RM02}.
 For example, the soft photons are often observed to
lag the high energy ones \citep{Cui99,V98} which is contrary to simple Comptonization
models, although recent studies have shown that under certain conditions a soft
lag could arise when more than one physical parameter is
oscillating \citep{LMT}.  Further the time lags for low frequency QPO, $ < 1$ Hz, are often
large, $> 0.1$ s \citep{Cui99} and hence are probably not due to any radiative
process and may be due to non-linear multiplicative reverberation effects \citep{Sha12}. Large time lags have also been observed for low frequency non periodic continuum fluctuations
\citep{Nowak} which are probably associated with
the slow propagation of waves in an accretion disk
\citep{Nowak,M00} or  correspond to the rise/decay time of magnetic
flares \citep{PF}.

 Perhaps the most unexpected results are that for some QPO,
the phase lag for the odd and even harmonics have opposite signs. 
Such alternating lag behavior
have been observed at different frequencies,
e.g. $67$ mHz \citep{Cui99} and $\approx 3$ Hz  \citep{Lin,Reig,TK} for
GRS 1915+105 and at $\approx 3$ Hz for XTE J1550-564 \citep{Cui00,Wijn}. 
Possible explanations for such  phenomena include partial covering models \citep{Var05}
or due to radiative transport of photons through a hot Comptonizing medium \citep{Bot00}. \citet{Ingram09} in their model proposed that QPO arises due to the Lense-Thirring precession of the hot inner flow. Based on this, \citet{Axe13} explained the average spectrum and
those of a QPO and its harmonic for XTE J1550-564, however it is not clear whether it can
also explain the complex time-lag as function of energy. 
 \cite{Axe13} quote a recent work by \cite{Vele13} where they have shown that the existance of a strong harmonic 
is predicted by the angular radiation pattern from comptonisation of a precessing hot inner flow. 
\citet{Bot00} show that alternating lags for the $\approx 3$ Hz 
 can be explained by time delays due to  Comptonization but such a model is not applicable for the lower
frequency QPO. Since alternating time lags for different harmonics have been observed for different frequencies
and for different systems, any model for the phenomena has to
be sufficiently generic and should not depend on the details
of QPO production and/or radiative mechanism.

In this work, such a generic model is presented.
It is shown that alternating lags will occur under fairly general
conditions if the QPO  is due to the oscillation of two dependent  
physical parameters. The two parameters could characterize a single
component of the spectrum or could correspond to two different spectral
components. Based on this interpretative framework, a phenomenological model is
developed which can explain the complex energy dependent features of
the $67$ mHz QPO observed in GRS 1915+105. In particular, the
amplitude and phase lags with energy for the fundamental and the next three
harmonics (a total of eight curves) is explained with this model consisting of
six parameters. Predictions are made for 
the energy dependence of the phases of the Bicoherence function which, in
principle, can be checked with further analysis of the observational data.

\section{Generic model for alternating lags} \label{sec: 2} 

Consider a photon spectrum $s(E)$, characterized by two physical parameters
$a$ and $b$, such that the temporal variations in the spectrum $s(E,t) 
\rightarrow s_o(E) +\Delta s (E,t)$, are due to corresponding variations
in $a(t) = a_o + \Delta a(t)$ and $b(t) = b_o + \Delta b(t)$. If
the response of the spectrum to the variations in the parameters is linear,
then
\begin{equation}
\Delta s(E,t)  = \gamma_a(E) \Delta a(t) + \gamma_b(E) \Delta b(t)
\label{eq:delS}
\end{equation}
where $\gamma_a$ and $\gamma_b$ are time independent functions of energy.
In the above equation it has been assumed
that the radiative process time-scale ( $\approx \tau L/c \approx 2$ msec, where $\tau \approx 1$ is
the optical depth and $L \approx 50GM/c^2 \approx 7 \times 10^7 cm$ is the size of the system for
a ten solar mass black hole) is much shorter than the variability
time-scale of the parameters. In other words, the time lag between the
parameter variations and the spectral response is negligible. 

In this scenario, the variations in the spectral parameters ($\Delta a(t)$ and 
$\Delta b(t)$) are caused by a driving dynamical oscillation which produces the QPO. 
It is further assumed that the variations in $b$ are also driven linearly by
variations in $a$ after a time lag $t_d$, that is,
\begin{equation}
\Delta b(t) = F \Delta a(t-t_d) + D(t)
\label{eq:delb_a}
\end{equation}

where $F$ is a time independent constant and $D(t)$ denotes the direct coupling between
the driving oscillation and $b$. 

Although the direct coupling term ($D(t)$) may not be negligible
compared to the interactive term ($F \Delta a(t-t_d)$) for some
systems ( e.g. section 3 below), it is
convenient to neglect $D(t)$ at this stage, only
to illustrate how such a model could produce alternating lags.With
this assumption a simple equation for the phase lags can be written
which explains the origin of the alternating lag behavior in a straight forward
manner. Neglecting $D(t)$ and
combining Eqn.
(\ref{eq:delS}) and (\ref{eq:delb_a}) gives,
\begin{equation}
\Delta S(E,\omega) = \gamma_a (E) \Delta A(\omega)[ 1 + \gamma_r (E) e^{-i\omega t_d}]
\label{eq:delSf}
\end{equation}
where, $S(E,\omega)$ and $A(\omega)$ are the Fourier transforms of
$s(E,t)$ and $a(t)$ respectively, $\omega$ is the angular frequency
and $\gamma_r (E) \equiv F \gamma_b (E)/\gamma_a (E)$. The
cross-spectrum $C(\omega,E_1,E_2)$ for two energies $E_1$ and $E_2$, is defined as
$\equiv \Delta S^*(E_1,\omega) \Delta S (E_2,\omega)$. Thus
\begin{eqnarray}
C (\omega,E_1,E_2) & = & \gamma_a(E_1) \gamma_a(E_2) | \Delta A(\omega)|^2 \times \nonumber \\ 
& & [1 +\gamma_r (E_1) e^{i\omega t_d}][1 + \gamma_r (E_2) e^{-i\omega t_d}]
\label{eq:cross}
\end{eqnarray}
The phase, $\phi(\omega,E_1,E_2)$, of the cross-spectrum divided by
$\omega$ is generally identified as the time lag between the photons
of energy $E_1$ and $E_2$. From the form of Eqn. \ref{eq:cross}, it
can be seen that this phase can be written as,
\begin{eqnarray}
\hbox {sin } \phi&  = & \frac{Im[C]}{|C|}\nonumber\\ &  = &{(\gamma_r (E_1) - \gamma_r (E_2)) \hbox {sin} (m \omega_o t_d) \over |1 +\gamma_r (E_1) e^{i\omega t_d}||1 + \gamma_r (E_2) e^{-i\omega t_d}|}
\label{eq:tphi}
\end{eqnarray}
Here $\omega$ has been replaced by $m \omega_o$ where $\omega_o$ is the
angular frequency of the fundamental oscillation (first harmonic) and different values
of the integer $m$ correspond to the other harmonics of the system. Since
the denominator of Eqn. \ref{eq:tphi} is always positive, the sign of
the phase angle is determined by $(\gamma_r (E_1) - \gamma_r (E_2)) \hbox {sin} (m \omega_o t_d)$. 
This implies that the system will exhibit alternating
lag signs for $n$ harmonics, provided $t_d$ lies within the range
\begin{equation}
\pi(1 -1/m) < \omega_o t_d < \pi (1 +1/m) 
\label{eq:cond}
\end{equation}
where $m = n+1$. For
example, for a system with four harmonics with alternating
lags (like the 67 mHz QPO in 
GRS 1915+105), the time lag $t_d$ is restricted to lie between
$3/8$ and $5/8$ of $1/f_o$ where $f_o$ is the fundamental frequency. Note
that the phase $\phi$ depends on function
$\gamma_r (E)$ (Eqn \ref{eq:tphi}) and in general could be small even
though $t_d$ is required to be $ \approx 1/f_o$ to satisfy condition (\ref{eq:cond}). 

It is useful to enumerate the different conditions that need to exist
in a system for this model to be applicable. The first condition is that
there needs to be at least two parameters which characterize the spectrum.
This is generally true for most radiative models invoked to explain hard
X-ray spectra. For example, in the Comptonization model, the spectrum depends
on the optical depth and temperature of the Comptonizing region. 
It should be noted that the spectral parameters $a$ and $b$ need not
necessarily characterize the same spectral component. For example,
$a$ could characterize a soft (black body like) component  of the  spectrum while $b$ could
be related to the hard X-ray power-law  which lags the intrinsic one
after a time delay.  The
second condition is that two of the parameters should be related to each other
(Eqn \ref{eq:delb_a}). Again it is probably true in general that
variations in one parameter will induce variations in the other. The third
condition is that $t_d$, the interaction time-scale between the two parameters,
is of the same order as $1/f_o$, where $f_o$ is the QPO frequency. This
would only be true for those phenomena where the QPO producing mechanism
is similar to the one coupling the two parameters. For example, if the
QPO is being produced by a dynamic process and the coupling between the
two parameters is also of a dynamic nature, then it is expected that
$t_d \approx 1/f_o \approx t_{dyn}$, where $t_{dyn}$ is the dynamic time-scale. 
Finally the fourth condition is the restriction on $t_d$ for the system
to show alternating lags (Eqn \ref{eq:cond}). If the previous
three conditions are satisfied for most of the systems, then this
final criterion is expected to occur in a fraction of them. It should
be emphasized that if there is an observation of alternating lags for
a large number of harmonics of a QPO, then for this
 model to be applicable, $t_d$ would need to be ``fine tuned'' to
satisfy Eqn \ref{eq:cond}. Thus such a observation would indicate
that this model is at least not in general applicable. However, for
alternating lags up to three or four harmonics, condition (\ref{eq:cond})
is not overtly restrictive. 
When the direct coupling $D(t)$
is not negligible, the criterion for a system to exhibit alternating
lags becomes more complex than Eqn. \ref{eq:cond} and depends on
the strength and form of the interaction. In such cases, as shown
in the next section, a more detailed calculation of the phase lags
is required.

\section{ Application to 67 millihertz QPO of GRS 1915+105} 

A 67 mHz QPO with the next three harmonics with alternating lags was detected 
by RXTE during 1996 May 5 observation of the black hole system GRS 1915+105
\citep{M97,Cui99}. Since the Q factor for this QPO is
large (FWHM $\approx$ mHz for the fundamental), the underlying continuum
is not expected to effect the  phase lag results. This and the fact that
a total of four harmonics were observed for this source makes this phenomenon
 an excellent candidate to test the generic model developed in the earlier
section. To compare the observed values with ones predicted from the model,
specific form and strength of the interactions between the driving
oscillation and the spectral parameters have to be chosen. Here, such
a specific set of interactions is presented which can explain the observed 
energy dependent trends of the QPO phase lag and amplitude.

Variations in the photon spectrum $s(E,t)$ are assumed to be linearly
driven by variations in two parameters $a$ and $b$ (Eqn \ref{eq:delS}), 
such that
\begin{equation}
{\Delta s(E,t) \over s_o(E)} = -E \Delta a (t) -\hbox {log} (E/E_p) \Delta b (t)
\label{eq:M1}
\end{equation}
where $s_o(E)$ is the time independent part of the spectrum 
The first term of the R.H.S is the linear response of the spectrum to
perturbation in $a$ if $s(E) \propto e^{-Ea}$, while the second
term is for the case when $s(E) \propto E^{-b}$. Thus Eqn
(\ref{eq:M1}) corresponds to the situation when the spectrum can
be described as an exponentially cut-off power-law ( $s(E) 
\propto E^{-b} e^{-Ea}$) with $a$ characterizing the inverse of 
the cutoff energy and $b$ the power-law slope. In this interpretation, $E{_p}$ is the pivot energy over which
power-law component varies.
This is partly motivated by observations of the high energy ($E > 3 keV$)
time averaged spectra of some black hole systems which can be approximately
described as an exponentially cut-off power law. The time averaged spectrum for the 
observation showing the 67 mHz QPO, can be roughly (at 2\% level) described as a
exponentially cut-off power-law of index  $\sim 1.8$ and cutoff energy $\sim 5$ keV.
However, in general detailed spectral analysis of GRS 1915+105 have shown that its spectrum is 
generally more complex \citep[e.g.][]{Yad99,Zdz01,Rod08}
To avoid such complexities and to keep the temporal
model presented here independent of specific spectral models, Eqn 
(\ref{eq:M1}) is used here in this analysis with the caveat that in
reality the response functions could be significantly more complex.
The motivation here is to show that the generic model developed in
section 2, is applicable to the 67 mHz QPO and there exists a
simple set of equations which can explain the complex behavior
of this phenomenon.

Since harmonics are observed in this system, at least one of
the parameters should couple non-linearly to the unknown driving
mechanism that produces the QPO. Further, since exactly four harmonics were observed, this suggests that the coupling is of a
quartic nature. Hence it is assumed here that,
\begin{equation}
\Delta a (t) = R_a (1 + \beta \hbox {cos}(\omega_o t))^4
\label{eq:M2}
\end{equation}
where $1 + \beta \hbox{cos}(\omega_o t)$ is the oscillating driver 
and $R_a$ is the coupling constant between it and $a$. These variations
in $a$ cause variations in $b$ (Eqn \ref{eq:delb_a}) with a time
delay $t_d$,

\begin{equation}
\Delta b(t) = F \Delta a(t-t_d) + R_b \beta \hbox{cos}(\omega_o t)
\label{eq:M3}
\end{equation}

\begin{figure}
\includegraphics[height=80mm, angle=0]{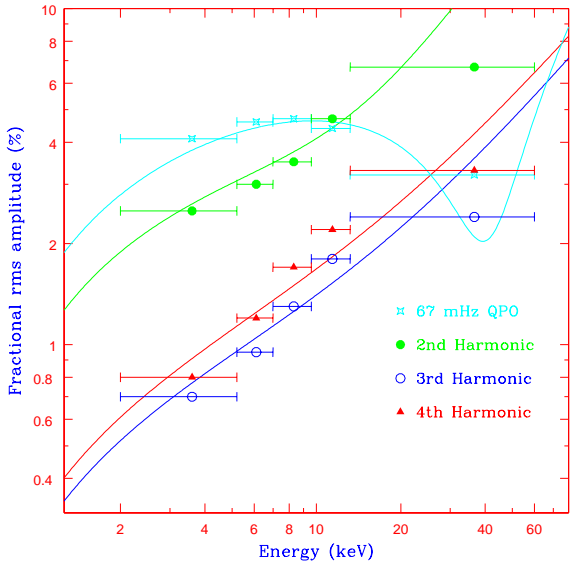}
\caption{Variation of the fractional rms amplitude with energy for
the 67 mHz QPO and the next three harmonics. Data points are from \citet{Cui99} and with errors
(not shown)  smaller than $0.2$ \%.  Solid
lines are model predictions (from top to bottom: fundamental (1st
harmonic), 2nd harmonic, 4th harmonic and 3rd harmonic). The six parameters used in this
fit are $E_p = 0.5$ keV, $R_a = 2.3 \times 10^{-6}$ keV$^{-1}$, $\beta = 9$,
$F = -4$ keV, $\omega_o t_d = 1.15 \pi$ and $R_b = -7.2\times 10^{-3}$. }
\label{fig1}
\end{figure}

\begin{figure}
\includegraphics[height=80mm, angle=0]{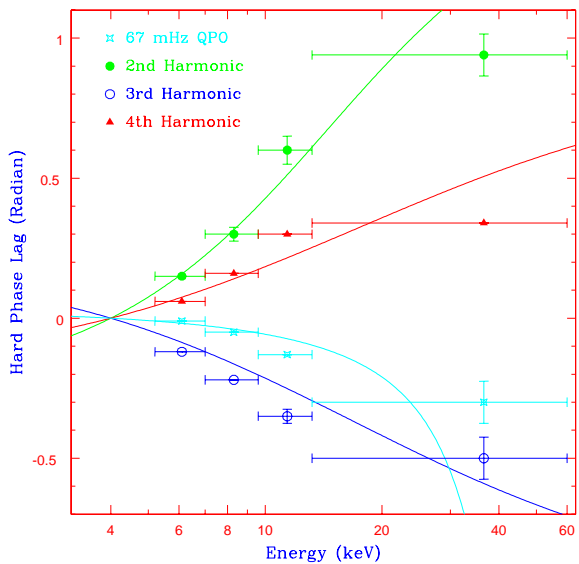}
\caption{Variation of the hard phase lag with energy for
the 67 mHz QPO and next three harmonics. Data points are from \citet{Cui99}. Points
without error bars have error smaller than the symbols.
Solid
lines are model predictions (from top to bottom:  2nd
harmonic, 4th harmonic, fundamental (1st harmonic)  and 3rd harmonic). The parameters used
for the model are same as in figure \ref{fig1}.}
\label{fig2}
\end{figure}

Here it is assumed that the direct coupling between the driving oscillation and $b$ ($D(t)$)
is linear and in phase with interaction causing $\Delta a(t)$.
Eqn (\ref{eq:M1}), (\ref{eq:M2}) and (\ref{eq:M3}) can
now be solved to obtain the fractional rms amplitude and the
phase lag as a function of energy for the fundamental and the next three
harmonics. A detailed fitting to the observed data points would
require convoluting the predicted time varying spectrum with the instrumental
response function and binning the resultant folded spectrum in appropriate energy
bins. However, since the motivation here is to show that qualitative trends
in the data can be explained by the model, the theoretical variations are directly
compared with the data points. The results are shown in 
figures \ref{fig1} and \ref{fig2}. The six parameters  required for the fit 
are $E_p$,$R_a$, $\beta$,$F$,$t_d$ and $R_b$. The phase lag as a function
of energy for the three harmonics other than fundamental (figure \ref{fig2}) depends only on three of these parameters,
$E_p$,$F$ and $t_d$. The variation of the fractional amplitude with energy 
of the same harmonics depends on two additional parameters $R_a$ and $\beta$,
while the amplitude and phase lag of the fundamental with energy also
depends on $R_b$.  

\begin{figure}
\includegraphics[height=80mm, angle=0]{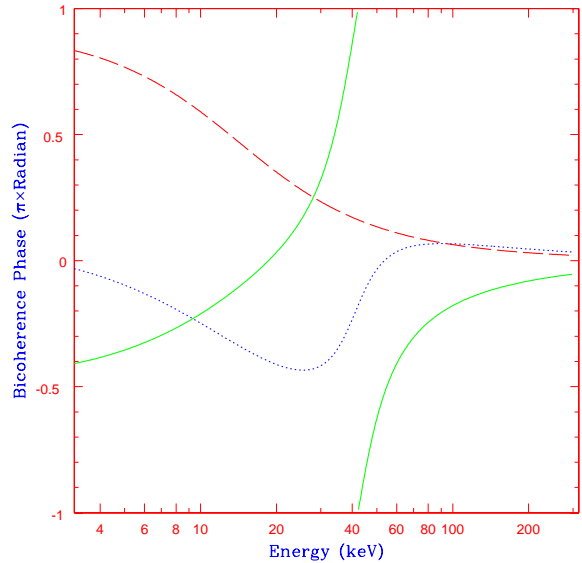}

\caption{Predicted variation of the phase of the Bicoherence
function with energy. The parameters for the model are same as in
figure \ref{fig1}. Solid line: phase of $B_{11}$ ; Dotted line: phase of $B_{12}$ ;
Dashed line: phase of $B_{22}$.}
\label{fig3}
\end{figure}

It is encouraging to find that the observed relative amplitude of the different
harmonics (including the observation that the amplitude of the fourth 
harmonic is larger than the third ) can be explained by a simple 
non-linear equation (\ref{eq:M2}). In particular the Fourier transform
of Eqn (\ref{eq:M2}) leads to
\begin{eqnarray}
{\Delta A(\omega)\over R_a} & =& (2\beta +{3\over 2}\beta^3)\delta (\omega -\omega_o)+ ({3\over 2}\beta^2 +{1\over 4}\beta^4)\delta (\omega -2\omega_o) \nonumber\\
& &
+ ({1\over 2}\beta^3)\delta (\omega -3\omega_o)
+ ({1\over 16}\beta^4)\delta (\omega -4\omega_o)
\end{eqnarray}
which when combined with Eqn (\ref{eq:M1}) and (\ref{eq:M3}) 
sets the relative amplitude of the different harmonics (figure \ref{fig1}).
However the interactions chosen here (Eqn. \ref{eq:M2},  \ref{eq:M3}) may not be unique 
and it may be possible
that there exists  more complex non-linear relationships which also
give similar results. For example, $a(t)$ could  be coupling  non-linearly to
$b(t)$ instead of the driver i.e. $a(t) = f_{nl}(b(t))$ , where $f_{nl}$ is
a nonlinear function.  However, such a function would give different phase
lags between the harmonics than those computed using Eqn (\ref{eq:M2}).
The phase lag between the harmonics can be indirectly measured using the
phases of the Bicoherence function which is defined as  (\citet{MC,MC11,MC13})
\begin{equation}
B_{nm}(E) \equiv \Delta S_n(E) \Delta S_m(E) \Delta S_{m+n}^*(E)
\label{eq:bicoher}
\end{equation}
where the integer sub
script on $\Delta S (E)$ stands for different harmonics.
Figure \ref{fig3} shows the predicted values of the phase of $B_{11}$, $B_{12}$ and
$B_{22}$ as a function of energy using Eqn (\ref{eq:M1}), (\ref{eq:M2})
and (\ref{eq:M3}). These variations can be directly compared with the
observed values to test the validity of these equations and confirm the
model developed here. Significant differences from the predicted values
would perhaps indicate that the non-linear equation (\ref{eq:M2}) is
more complex than the one assumed in this work.


\section{ Summary and discussion} 

The phenomenological model presented here should be developed further
to incorporate the specific radiative and dynamic mechanism that
could be  operating for this system. 
The inferred weak coupling between the parameters($a$ and $b$) and the 
driving oscillation  
($R_a =  2.3\times 10^{-6}$ keV$^{-1}$ and $R_b = -7.2 \times 10^{-3}$), could
occur 
if the driving oscillation is localized to some particular radius
of the accretion disk and hence couples weakly with the parameters
which are probably  global averages. A localized driver would also
naturally explain why the QPO is so narrow ( FWHM $\approx$ mHz) since
then the frequency could be associated with some natural frequency of
the disk at that particular radius. However,
it would then be difficult to understand why the time scale of interaction
between the two (global) parameters $t_d$ is similar in magnitude
to the timescale of the oscillation ($1/f$). If the form of the equation
(\ref{eq:M2}) is correct that would indicate that the physical
quantity associated with the driver can have negative values since
$\beta = 9 > 1$. This means that the driving oscillation is not
a positive physical quantity like temperature, density etc but 
needs to be associated with a quantity which can have
negative values e.g. accretion rate, viscous stress etc.  

In summary, a generic model for alternating lags for the
harmonics of QPO has been presented. Based on the generic
model, a phenomenological one has been developed
to explain the amplitude and phase lag variation with energy of
the harmonics of the 67 mHz QPO observed in GRS 1915+105.
The model can be further strengthened and/or modified by comparing
the predicted values of the phase of the Bi-coherence function with
the ones inferred from observations. The model also puts restrictions
on the radiative and dynamic mechanisms that are operating to produce
this phenomenon. Physical models developed in the framework of
the phenomenological one will enhance our understanding of these
systems.

\section{Acknowledgements}
The research leading to these results has been partially funded by ISRO-RESPOND program. SM gratefully acknowledges IUCAA for the visiting associateship.


\begin{thebibliography}{99}


\bibitem[Axelsson et al.(2013)]{Axe13} Axelsson, M., Done, C.,
\& Hjalmarsdotter,L.\ 2013, arXiv:1307.4396

\bibitem[B\"ottcher \& Liang (2000)]{Bot00} B\"ottcher, M., \& Liang, E.P. \ 2000, arXiv:astro-ph/0003139

\bibitem[Cui(1999)]{Cui99} Cui, W.\ 1999, \apjl, 524, L59 

\bibitem[Cui et al.(2000)]{Cui00} Cui, W., Zhang, S.~N., 
\& Chen, W.\ 2000, \apjl, 531, L45 


\bibitem[Ingram et al.(2009)]{Ingram09} Ingram, A., Done, C., 
\& Fragile, P.~C.\ 2009, \mnras, 397, L101 


\bibitem[Lee et al.(2001)]{LMT} Lee, H.~C., Misra, R., 
\& Taam, R.~E.\ 2001, \apjl, 549, L229 

\bibitem[Li et al.(2013)]{Li13} Li, Z.~B., Qu, J.~L., Song, 
L.~M., Ding, G.~Q., \& Zhang, C.~M.\ 2013, \mnras, 428, 1704



\bibitem[Lin et al.(2000)]{Lin} Lin, D., Smith, I.~A., 
Liang, E.~P., \& B\"ottcher, M.\ 2000, \apjl, 543, L141 



\bibitem[Maccarone \& Coppi(2002)]{MC} Maccarone, T.~J., \& Coppi, P.~S.\ 2002, \mnras, 336, 817 

\bibitem[Maccarone et al.(2011)]{MC11} Maccarone, T.~J.,Uttley, P., van der Klis, M., Wijnands, R.~A.~D.,
 \& Coppi, P.~S.\ 2011, \mnras, 413, 1819


\bibitem[Maccarone(2013)]{MC13} Maccarone, T.~J.\ 2013, 
\mnras, 2181 


\bibitem[McClintock \& Remillard (2006)]{MR06} McClintock J. E. \& Remillard R. A., \ 2006 , Compact Stellar X-ray Sources , Cambridge University Press, p.157

\bibitem[Miller et al.(1998)]{Mi98} Miller, M.~C., Lamb, 
F.~K., \& Psaltis, D.\ 1998, \apj, 508, 791 

\bibitem[Misra(2000)]{M00} Misra, R.\ 2000, \apjl, 529, L95 





\bibitem[Morgan et al.(1997)]{M97} Morgan, E.~H., 
Remillard, R.~A., \& Greiner, J.\ 1997, \apj, 482, 993 

\bibitem[Nowak et al.(1999)]{Nowak} Nowak, M.~A., Vaughan, 
B.~A., Wilms, J., Dove, J.~B., \& Begelman, M.~C.\ 1999, \apj, 510, 874


\bibitem[Poutanen 
\& Fabian(1999)]{PF} Poutanen, J., \& Fabian, A.~C.\ 1999, \mnras, 306, L31 

\bibitem[Qu et al.(2010)]{QU10} Qu, J.~L., Lu, F.~J., Lu, 
Y., et al.\ 2010, \apj, 710, 836 


\bibitem[Reig et al.(2000)]{Reig} Reig, P., Belloni, T., van 
der Klis, M., et al.\ 2000, \apj, 541, 883 

\bibitem[Remillard et al.(2002)]{RM02} Remillard, R.~A., 
Sobczak, G.~J., Muno, M.~P., \& McClintock, J.~E.\ 2002, \apj, 564, 962 

\bibitem[Remillard 
\& McClintock(2006)]{RM06} Remillard, R.~A., \& McClintock, J.~E.\ 2006, \araa, 44, 49 


\bibitem[Rodriguez et 
al.(2002)]{RJ02} Rodriguez, J., Durouchoux, P., Mirabel, I.~F., et al.\ 2002, \aap, 386, 271



\bibitem[Rodriguez et al.(2008)]{Rod08} Rodriguez, J., Shaw, S. E., Hannikainen, D. C., Belloni, T., Corbel, S., Cadolle Bel, M., Chenevez, J., Prat, L., et al., 2008, ApJ, 675, 1449

\bibitem[Shaposhnikov(2012)]{Sha12} Shaposhnikov, N.\ 2012, 
\apjl, 752, L25 


\bibitem[Stella 
\& Vietri(1999)]{SV99} Stella, L., \& Vietri, M.\ 1999, Physical Review Letters, 82, 17

\bibitem[Titarchuk 
\& Osherovich(1999)]{T099} Titarchuk, L., \& Osherovich, V.\ 1999, \apjl, 518, L95 


\bibitem[Tomsick 
\& Kaaret(2001)]{TK} Tomsick, J.~A., \& Kaaret, P.\ 2001, \apj, 548, 401


\bibitem[Varni{\`e}re(2005)]{Var05} Varni{\`e}re, P.\ 2005, \aap, 434, L5 




\bibitem[van der 
Klis(2000)]{V00} van der Klis, M.\ 2000, \araa, 38, 717

\bibitem[van der 
Klis(2006)]{V06} van der Klis, M.\ 2006, Compact Stellar X-ray Sources , Cambridge University Press, p.39

 
\bibitem[Vaughan et al.(1998)]{V98} Vaughan, B.~A., van der 
Klis, M., M{\'e}ndez, M., et al.\ 1998, \apjl, 509, L145


\bibitem[Veledina et al.(2013)]{Vele13} Veledina, A., 
Poutanen, J., \& Ingram, A.\ 2013, MNRAS submitted. 



\bibitem[Wijnands et al.(1999)]{Wijn} Wijnands, R., Homan, 
J., \& van der Klis, M.\ 1999, \apjl, 526, L33


\bibitem[Yadav et al.(1999)]{Yad99} Yadav, J. S., Rao, A. R., Agrawal, P. C., Paul, B., Seetha, S. \& Kasturirangan, K. 1999, ApJ, 517, 935


\bibitem[Zdziarski et al.(2001)]{Zdz01} Zdziarski, A. A., Grove, J. E., Poutanen, J., Rao, A. R., \& Vadawale, S. V. 2001, ApJ, 554, L45


\end{thebibliography}
\end{document}